\documentclass[12pt]{article}
\pdfoutput=1
\usepackage{dcolumn}
\usepackage{bm}
\usepackage{graphicx}
\usepackage{amssymb,amsmath}
\usepackage{multirow}
\usepackage{afterpage}
\usepackage{cite,color,url}
\usepackage[colorlinks=true
,urlcolor=magenta
,anchorcolor=blue
,citecolor=blue
,filecolor=blue
,linkcolor=blue
,menucolor=blue
,linktocpage=true
,pdfproducer=medialab
,pdfa=true
]{hyperref}
\usepackage{cleveref}
\usepackage{tikz-feynman}
\usetikzlibrary{trees}
\usetikzlibrary{decorations.pathmorphing}
\usetikzlibrary{decorations.markings}

\usepackage{array,booktabs}
\usepackage{subcaption}

\usepackage{slashed}
\usepackage{epsfig,psfrag,rotating,soul}
\usepackage{rotfloat}
\usepackage[font={small}]{caption}
\usepackage{colortbl}
\usepackage{array,booktabs}
\usepackage{subcaption}
\usepackage{contour}
\usepackage[normalem]{ulem}

\evensidemargin \oddsidemargin
\allowdisplaybreaks

\let\OLDthebibliography\thebibliography
\renewcommand\thebibliography[1]{
  \OLDthebibliography{#1}
  \setlength{\parskip}{0pt}
  \setlength{\itemsep}{0pt plus 0.3ex}
}

\newcommand{\beq}{\begin{equation}}
\newcommand{\eeq}{\end{equation}}

\usepackage{xspace}

\newcommand{\ba}{\begin{array}}
	\newcommand{\ea}{\end{array}}
\newcommand{\bd}{\begin{displaymath}}
	\newcommand{\ed}{\end{displaymath}}
\newcommand{\besub}{\begin{subequations}}
	\newcommand{\eesub}{\end{subequations}}
\newcommand{\bea}{\begin{eqnarray}}
	\newcommand{\eea}{\end{eqnarray}}

\def\q2 {q^2}

\def\bt{\begin{table}}
\def\et{\end{table}}

\definecolor{mygray}{gray}{0.85} 
\definecolor{myblue}{cmyk}{0.65, 0.37, 0.0, 0.19}

\begin{document}
\thispagestyle{empty}

\def\thefootnote{\fnsymbol{footnote}}

\vspace*{1cm}

\begin{center}

\begin{Large}
\textbf{\textsc{ Neutron stars as thermometers for reheating induced dipole dark matter
}}
\end{Large}

\vspace{0.5cm}
 Sahabub ~Jahedi$^{1,2}$%
\footnote{{\tt \href{mailto:sahabub@m.scnu.edu.cn}{sahabub@m.scnu.edu.cn}}}%

\vspace*{.2cm}

{\it
$^1$State Key Laboratory of Nuclear Physics and Technology, Institute of Quantum Matter, South China Normal 
	University, Guangzhou 510006, China\\
	
$^2$Guangdong Basic Research Center of Excellence for Structure and Fundamental Interactions of Matter, Guangdong 
	Provincial Key Laboratory of Nuclear Science, Guangzhou 510006, China\\
}

\end{center}

\vspace{0.1cm}

\begin{abstract}

We investigate the electromagnetic interactions of dipole dark matter (DM) within an effective field theory framework, considering both standard and non-standard cosmological scenarios. We first study the prospects of DM production via both the freeze-out and freeze-in mechanisms within the standard radiation-domination. We then investigate how the viable DM parameter space is modified in a non-standard cosmological scenario due to entropy dilution during reheating. Existing constraints on the parameter space are discussed, and we highlight the discovery potential of future direct detection experiments to probe these scenarios. We further investigate the implications of neutron star heating for dipole DM. Due to the momentum-dependent nature of the interaction, dipole DM is captured efficiently by neutron stars, thereby making neutron star heating a sensitive probe of the dipole DM parameter space.

\end{abstract}

\def\thefootnote{\arabic{footnote}}
\setcounter{page}{0}
\setcounter{footnote}{0}

\newpage

\section{Introduction}
\label{sec:intro}
Although the astrophysical and cosmological evidences for dark matter (DM) are highly compelling \cite{Zwicky:1933gu,Rubin:1980zd,Schlegel:1997yv}, its fundamental nature remains an open question. A wide range of possibilities for DM candidate has been considered, spanning many orders of magnitude in mass and coupling strength \cite{Bertone:2004pz,Cirelli:2024ssz}. One of the most critical criteria for evaluating the viability of a DM model is that it predicts relic density measured by the Planck \cite{Planck:2018vyg}. The calculation of the DM relic density, whether through the freeze-out \cite{Lee:1977ua,Gondolo:1990dk,Jungman:1995df} or freeze-in \cite{Hall:2009bx,Elahi:2014fsa,Bernal:2017kxu} mechanisms, is typically performed within the standard cosmological scenario, which assumes that the energy density of the Universe was dominated by the Standard Model (SM) radiation from the end of inflationary reheating until the onset of Big Bang Nucleosynthesis (BBN). Within this standard scenario, the reheating temperature ($T_{\text{RH}}$), defined as the beginning of the radiation-dominated era, is larger than the temperature at which the DM is produced. The inflaton, a scalar field, decays into SM radiation during reheating, leads to an increase in the entropy of the Universe throughout this phase \cite{Allahverdi:2020bys,Batell:2024dsi}. The decay width of the inflaton also determines the value of $T_{\text{RH}}$ and the maximum temperature ($T_{\text{max}}$) of the universe. The $T_{\text{RH}}$ is not strongly constrained, and consistency with the successful predictions of BBN allows it to take any value above a few MeV \cite{Sarkar:1995dd, Kawasaki:2000en, Hannestad:2004px, DeBernardis:2008zz, deSalas:2015glj}.

Injection of entropy during the reheating can significantly relax the required thermally averaged annihilation cross-section for weakly interacting massive particle (WIMP), potentially lowering it by several orders of magnitude. As a result, WIMPs with weaker couplings to the SM would evade the current direct search constraints. In the context of simplified models and effective field theories, several studies have demonstrated that departures from the standard cosmological framework can substantially alter the predicted DM relic abundance, for example via modifications to the early-universe entropy evolution during reheating for WIMPs \cite{Bernal:2022wck,Bernal:2024yhu,Bhattacharya:2025wef,Mitra:2025cmo,Mondal:2025awq,Pradhan:2026maz}. On the other hand, in scenarios with FIMP, entropy injection enhances the efficiency of DM production through the freeze-in mechanism. Consequently, a larger production cross-section, or equivalently, a stronger coupling, is required to reproduce the observed relic abundance. In this direction, various studies have found that if the reheating temperature is taken below the typical freeze-in scale ($T_{\text{FI}} \sim m_{\text{DM}}$), where $m_{\text{DM}}$ denotes the DM mass responsible for producing FIMPs, the couplings needed to obtain the observed relic density become correspondingly larger \cite{Mondal:2025awq,Pradhan:2026maz,Barman:2020plp,Barman:2021tgt,Bhattiprolu:2022sdd,Cosme:2023xpa,Boddy:2024vgt,Barman:2024nhr,Barman:2024tjt,Ghosh:2024nkj,Bernal:2024ndy,Belanger:2024yoj,Bernal:2025osg,Borah:2025ema,Khan:2025dht,DEramo:2025fvy,Drewes:2026uor,Endo:2026hsh}. In both scenarios, $T_{\text{RH}}$ significantly influences the DM relic density, in addition to the DM mass and its couplings. This, in turn, can improve the prospects for detection in future direct search experiments.

 In this work, we employ effective electromagnetic dipole interactions \cite{Sigurdson:2004zp,Masso:2009mu} to examine DM production via both freeze-out and freeze-in mechanisms during and after the reheating phase. We study existing direct, indirect, and collider constraints, and explore the possibility of probing the relic-allowed parameter space via future direct search experiments and neutron star heating. When a compact object moves through the DM halo, DM particles can scatter off SM constituents inside the object and lose sufficient energy to become gravitationally bound. The subsequent accumulation and annihilation or thermalization of DM can inject energy into the system, leading to observable heating of compact objects such as neutron star\footnote{Several phenomenological studies of neutron star heating and its implications for DM searches can be found in Refs.~\cite{Baryakhtar:2017dbj,Joglekar:2019vzy,Maity:2021fxw,Bose:2023yll}.}. This induced heating may produce detectable infrared signals, potentially observable with next-generation telescopes like the James Webb Space Telescope (JWST) \cite{Gardner:2006ky}. Furthermore, in the vicinity of neutron star, DM particles can be accelerated to velocities significantly larger than the typical galactic halo value, making relativistic effects important in the scattering and capture processes. This is in contrast to conventional direct detection experiments, where the non-relativistic approximation is generally valid.

This paper is organized as follows. In section~\ref{sec:reheating}, we briefly review the cosmological background equations relevant for reheating. In section~\ref{sec:pheno}, we introduce the effective electromagnetic dipole interactions of DM within an effective field theory framework. DM production during and after reheating is discussed in section~\ref{sec:dm_prod}. In section~\ref{sec:cons}, we examine the existing constraints from direct detection, indirect searches, and collider experiments. Section~\ref{sec:ns_heating} demonstrates neutron star heating in the context of dipole DM. Finally, we summarize our results in section~\ref{sec:results} and conclude in section~\ref{sec:conclude}.
\section{Reheating dynamics}
\label{sec:reheating}
During the reheating phase of the early Universe, the inflaton field, $\phi$, decays into SM radiation with a total decay width $\Gamma_\phi$. The evolution of the cosmological background is governed by a set of Boltzmann equations that describe the inflaton energy density, $\rho_\phi$, and the SM radiation energy density, $\rho_R$, \cite{Gelmini:2006pw}
\begin{align}
	\begin{split}
		\frac{d\rho_{\phi}}{dt} + 3\mathcal{H} \rho_{\phi} &= - \Gamma_{\phi} \rho_{\phi},\\
		\frac{d\rho_R}{dt} + 4\mathcal{H}\rho_R &= \Gamma_{\phi} \rho_{\phi}.
	\end{split}
	\label{eq:inf_rad_v1}     
\end{align}
The evolution of $\rho_R$ can also be reduced to
\begin{equation}
	\frac{ds}{dt} + 3\mathcal{H}s = \frac{\Gamma_{\phi} \rho_{\phi}}{T},
\end{equation}
where the entropy density ($s$) is defined as a function of SM radiation temperature, $T$,
\begin{equation}
	s=\frac{2\pi^2}{45}h_{\text{eff}}(T)T^3.
\end{equation} 
Here, $h_{\text{eff}}(T)$ denotes the effective number of relativistic degrees of freedom contributing to the entropy density~\cite{Drees:2015exa}. The Friedmann equation, which sets the Hubble expansion rate, $\mathcal{H}$, is given by
\begin{equation}
	\mathcal{H}^2 = \frac{8\pi}{3} \frac{\rho_\phi + \rho_R}{M_P^2},
	\label{eq:hub_eq}
\end{equation}
with $M_P \simeq 1.2 \times 10^{19}$~GeV denotes the Planck mass. The energy density of the SM radiation bath, $\rho_R$, is written as
\begin{equation}
	\rho_R(T) = \frac{\pi^2}{30} g_{\text{eff}}(T) T^4,
	\label{eq:rho_red}
\end{equation}
where $g_{\text{eff}}(T)$ represents the effective number of relativistic degrees of freedom contributing to $\rho_R$. For numerical evaluation, it is convenient to parametrize eq.~\eqref{eq:inf_rad} in terms of comoving variables $z_{\phi}=\rho_\phi \times a^3$ and $z_{R}=\rho_R \times a^4$, where $a$ is the scale factor. These variables satisfy
\begin{align}
	\begin{split}
		\frac{dz_{\phi}}{da} &= - \frac{\Gamma_{\phi}}{\mathcal{H}a}z_{\phi},\\
		\frac{dz_{R}}{da} &=  \frac{\Gamma_{\phi}}{\mathcal{H}}z_{\phi}.
	\end{split}
	\label{eq:inf_rad}     
\end{align}

\begin{figure}[t]
	\centering
	\includegraphics[width=0.49\linewidth]{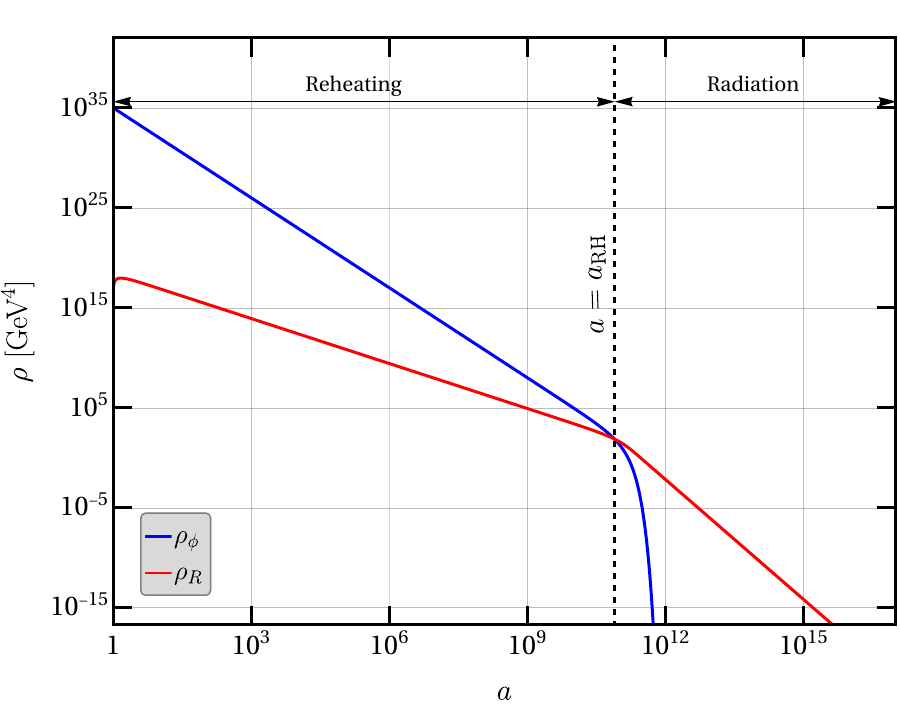}~~
	\includegraphics[width=0.49\linewidth]{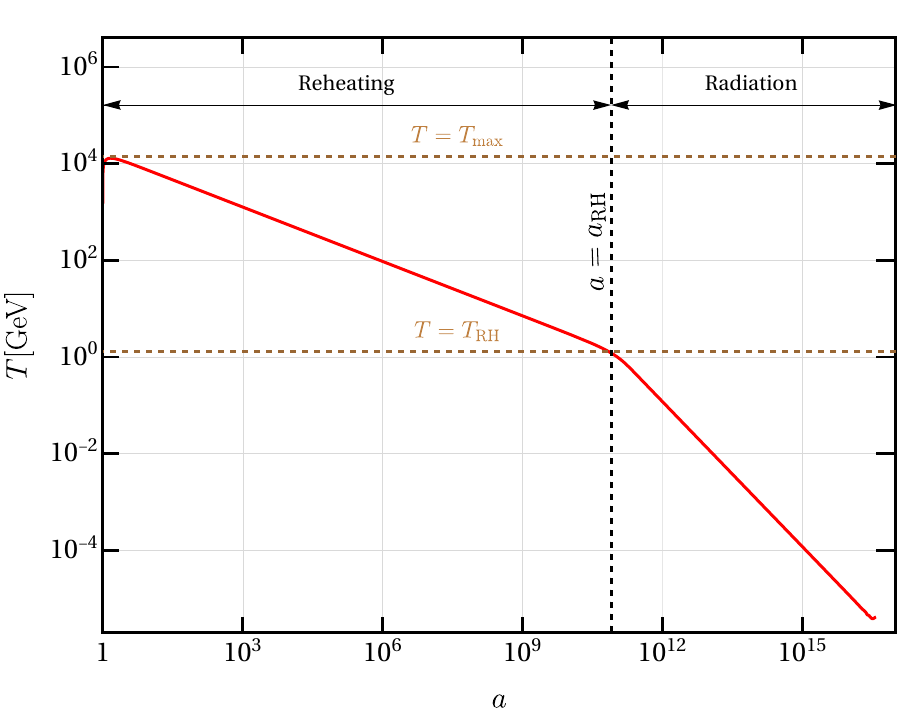}
	\caption{Left: evolution of inflation and radiation energy densities with the scale factor a. Right: evolution of SM temperature with the scale factor a. The vertical line in the both plots denote the separation between the reheating and radiation era.}
	\label{fig:inf_rad}
\end{figure}

To numerically solve eq.~\eqref{eq:inf_rad}, we impose the initial conditions at the onset of reheating, $t=t_I$, corresponding to the normalized scale factor $a_I \equiv a(t_I)=1$. At this stage, the radiation energy density vanishes, $\rho_R(t_I)=0$, while the inflaton energy density is given by $\rho_\phi(t_I)=\frac{3}{8\pi} M_P^2 \mathcal{H}_I^2$ with $\mathcal{H}_I$ denoting the Hubble scale at the end of inflation. Current limits from the non-observation of primordial $B$-modes constrain the inflationary scale to be $H_I < 4.0 \times 10^{-6} M_P$~\cite{BICEP:2021xfz}. In left panel of figure~\ref{fig:inf_rad}, we present the numerical evolution of $\rho_\phi$ and $\rho_R$ as functions of the scale factor ($a$) for the benchmark choice $H_I = 0.1$ GeV and $\Gamma_\phi = 10^{-17}\,\mathrm{GeV}$, corresponding to a reheating temperature of $T_{\text{RH}} \simeq 1.8$ GeV. During the reheating epoch, for the quadratic nature of the inflaton potential, the inflaton energy density varies as $\rho_\phi(a)\propto a^{-3}$, whereas the radiation energy density evolves as $\rho_R(a)\propto a^{-3/2}$. Once radiation domination begins, the inflaton density rapidly becomes negligible due to exponential decay, and the standard radiation scaling $\rho_R(a)\propto a^{-4}$ is recovered. The evolution of the SM plasma temperature is displayed in the right panel of figure~\ref{fig:inf_rad}. Since the radiation bath is continuously sourced by inflaton decays during reheating, the temperature scales as $T(a)\propto a^{-3/8}$. After reheating is completed and entropy production ceases, the usual radiation-dominated behavior, $T(a)\propto a^{-1}$, is restored.

The conclusion of reheating is defined as the onset of SM radiation domination, at a temperature $T = T_{\text{RH}}$ defined as $\rho_{\phi}(T_{\text{RH}}) = \rho_{\phi}(T_{R})$. To maintain consistency with the successful predictions of BBN, one must require $T_{\text{rh}} > T_{\text{BBN}} \simeq 4$~MeV~\cite{Sarkar:1995dd, Kawasaki:2000en, Hannestad:2004px, DeBernardis:2008zz, deSalas:2015glj}. An alternative estimate for the reheating temperature follows from imposing $\mathcal{H}(T_{\text{rh}}) = \Gamma_\phi$, which yields
\begin{equation}
	T_{\text{RH}}^2 = \frac{3}{2\pi} \sqrt{\frac{5}{\pi\, g_s(T_{\text{rh}})}} M_P \Gamma_\phi.
\end{equation}
Furthermore, the maximum temperature $T_{\text{max}}$ attained by the thermal bath during reheating can be approximated as~\cite{Barman:2021ugy}
\begin{equation}
	T_{\text{max}}^4 = \frac{15}{2 \pi^3\, g_s(T_{\text{max}})} \left(\frac38\right)^{8/5} M_P^2\, \Gamma_\phi\, \mathcal{H}_I\,.
\end{equation}
\section{Phenomenological framework}
\label{sec:pheno}
Since a Majorana fermion cannot possess a permanent electromagnetic dipole moment due to its self-conjugate nature, we consider the DM candidate to be a Dirac fermion, denoted by $\chi$. We assume that $\chi$ is a singlet under the SM gauge symmetry. A discrete $\mathbb{Z}_2$ symmetry is imposed, under which $\chi$ is odd while all SM fields are even, ensuring the stability of the DM particle and fermion bilinear term. At the effective field theory (EFT) level, the leading interaction between the DM and the photon arises from dimension-5 electromagnetic dipole operators, given by \cite{Sigurdson:2004zp}
\beq
\mathcal{L}_{\text{EFT}}=\frac{1}{\Lambda}\bar{\chi}\sigma^{\mu \nu}(C_M + i\gamma5C_E) \chi F_{\mu \nu},
\label{eq:dipole_lag}
\eeq 
where $\Lambda$ is the cut-off scale of the EFT, $C_{M(E)}$ is the dimensionless Wilson coefficient (WC) of the magnetic (electric) dipole interaction, $\sigma^{\mu \nu}=\frac{i}{2}[\gamma^{\mu},\gamma^{\nu}]$ is the Lorentz generator in the spinor representation, and $F_{\mu \nu}=(\partial_{\mu}A_{\nu}-\partial_{\nu}A_{\mu})$ is the $U(1)_{\text{em}}$ field strength tensor. Throughout the analysis, we fix the WCs by setting $C_{M/E} = 1$ without loss of generality. Such interactions can be generated by integrating out a heavy scalar and vector like lepton contributing through one-loop diagram. A UV-complete model, along with the matching to the magnetic dipole interaction, is discussed in Appendix~\ref{sec:uv_completion}.  To keep the discussion as general as possible, we consider two representative scenarios: purely magnetic dipole DM ($C_E = 0$) and purely electric dipole DM ($C_M = 0$). These dipole interactions enable the production of DM in the early Universe and play a crucial role in establishing the thermal population of the dark sector, primarily via $s$-channel SM charged fermions annihilation and di-photon $t$-channel annihilation. In this work, we examine the phenomenology of both freeze-out and freeze-in scenarios within two distinct cosmological frameworks. The first is the conventional post-reheating picture, where DM decouples in radiation domination epoch. The second is an alternative setup in which the freeze-out or freeze-in takes place during the reheating epoch, within the non-standard cosmological timeline. Therefore, in the latter case, the DM abundance is influenced by the decay of the inflaton and, consequently, by the reheating temperature of the universe. These dipole interactions contribute to direct detection through DM scattering off nuclei and electrons, to indirect detection via annihilation into photons and charged fermions, and to collider searches through missing energy signatures associated with visible particles. They can also induce heating of neutron stars through the capture and subsequent trapping of DM in the dense stellar environment. We first discuss constraints from existing upper bounds from direct detection, indirect searches, and collider experiments, and then explore the prospects for probing these interactions via neutron star heating.

\section{Dark matter production}
\label{sec:dm_prod}
Depending on the strength of the DM-SM interaction cross-section, DM can be produced through either thermal processes (freeze-out) or non-thermal processes (freeze-in). In this section, we discuss the freeze-out and freeze-in production mechanisms of DM mediated by electromagnetic dipole interactions. We analyze both scenarios, i.e., during the radiation-dominated epoch as well as throughout the reheating phase.
\subsection{Freeze-out}
\label{sec:fo}

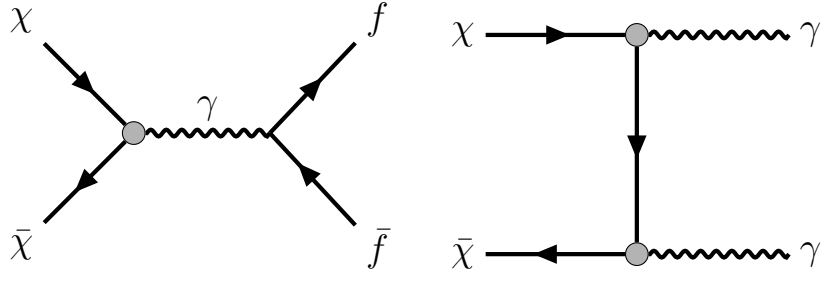
\begin{figure}[htb!]
	\centering
	\begin{tikzpicture}
		\begin{feynman}
			\vertex [blob, fill=gray!60, minimum size=0.3cm](a){};
			\vertex [above left=2.1cm of a] (b) {\large $\chi$};
			\vertex [below left=2.1cm of a] (c) {\large $\bar{\chi}$};
			\vertex [right=1.8cm of a] (d);
			\vertex [above right=1.6cm of d] (e) {\large $f$};
			\vertex [below right=1.6cm of d] (f) {\large $\bar{f}$};
			\diagram{
				(b) -- [fermion, ultra thick, arrow size=2pt] (a) -- [ fermion, ultra thick, arrow size=2pt] (c);
				(a) -- [ ultra thick, boson, edge label={\large $\gamma$}] (d);
				(d) -- [fermion, ultra thick, arrow size=2pt] (e);
				(d) -- [anti fermion, ultra thick, arrow size=2pt] (f);
			};
		\end{feynman}
	\end{tikzpicture}~~~
	\begin{tikzpicture}
		\begin{feynman}
			\vertex [left=2.3cm of a] (b) {\large $\chi$};
			\vertex [right=2.3cm of a] (c) {\large $\gamma$};
			\vertex [blob, fill=gray!60, minimum size=0.3cm](a){};
			\vertex [blob, fill=gray!60, minimum size=0.3cm, below=2.9cm of a](d){};
			\vertex [left=2.3cm of d] (e) {\large $\bar{\chi}$};
			\vertex [right=2.3cm of d] (f) {\large $\gamma$};
			
			\diagram* {
				(b) -- [fermion, ultra thick, arrow size=2pt] (a) -- [boson, ultra thick] (c);
				(a) -- [fermion, ultra thick, arrow size=2pt] (d);
				(e) -- [anti fermion, ultra thick, arrow size=2pt] (d) -- [boson, ultra thick, arrow size=2pt] (f);
			};
		\end{feynman}
	\end{tikzpicture}
	\caption{Feynman diagrams that contributes to the DM relic density. $f$'s are the SM fermions. The gray blob represents the EFT vertex.}
	\label{fig:feymann_diag}
\end{figure}

In the context of thermal production, commonly referred to as the freeze-out mechanism, DM is initially in thermal equilibrium with the SM bath. As the Universe expands and cools down, the DM interaction rate eventually drops below the $\mathcal{H}$. At this point, DM can no longer maintain chemical equilibrium with the thermal bath and decouples from it. Consequently, the comoving number density of DM becomes effectively constant. The evolution of the DM number density throughout this process is governed by the Boltzmann equation
\beq
\frac{dn_{\chi}^{}}{dt}+3\mathcal{H}n_{\phi}^{}=-\langle\sigma v\rangle(n^{2}_{\chi}-n^{2}_{\rm eq}),
\label{eq:beq}
\eeq
with $n_{\rm eq}\simeq g_{\chi}(m_{\chi}^{}~T/2\pi)^{3/2}e^{-m_{\chi}^{}/T}$ is the DM equilibrium number density, where $g_{\chi}$ and $m_{\chi}^{}$ are the number of degrees of freedom and mass of DM, respectively. During radiation domination, since the SM entropy is conserved, therefore, eq.~\eqref{eq:beq} can be written in terms of a dimensionless quantity $x=m_{\chi}/T$ and DM yield $Y_{\chi}=n_{\chi}/s$ as
\beq
\frac{dY_{\chi}}{dx}=-\frac{s\langle \sigma v \rangle}{\mathcal{H}x}(Y^2-Y^2_{\text{eq}}),
\eeq
where $s=(2\pi^2/25)g_s(T) T^2$ is the SM entropy density and $g_s(T)$ is the degrees of freedom of the thermal bath associated with $s$. The $\mathcal{H}$ is expressed as $\mathcal{H}=\sqrt{8 \pi \rho_R/3M_{\text{Pl}}^2}$ with $\rho_R$ defined in eq.~\eqref{eq:rho_red}.
 The quantity $\langle\sigma v\rangle$ denotes the thermal average cross section times relative velocity, defined as~\cite{Gondolo:1990dk}
\beq
	\langle\sigma v\rangle=\frac{g^2~T}{32\pi^4~n_{\rm eq}^{2}}\int_{4m_{\chi}^{2}}^{\infty}\sqrt{s}(s-4m_{\chi}^{2})\sigma(s)K_{1}\left(\frac{\sqrt{s}}{T}\right)~ds,
\label{eq:sigv}
\eeq
where $\sigma(s)$ is the DM annihilation cross-section to SM. The annihilation channels contributing to the relic density include the $s$-channel process $\chi \bar{\chi} \to f \bar{f}$ and the $t$-channel process $\chi \bar{\chi} \to \gamma \gamma$, as illustrated in figure~\ref{fig:feymann_diag}.
The analytical expressions for the $s$-channel $\chi \bar{\chi} \to f \bar{f}$ annihilation cross sections are given by
\bea
\begin{split}
&\sigma_{ff}=N_f\frac{2\alpha_e}{3\Lambda^2}(1-4\beta_f)\\& \times
\begin{cases}
\left[1-4\beta_f(1-2\beta_\chi+16\beta_\chi^2)-8\beta_f^2(1-4\beta_\chi)^2+10\beta_\chi-8\beta_\chi^2\right]; ~\text{magnetic},\\
(1-4\beta_\chi)\left[1+4\beta_f(1-2\beta_f)(1-4\beta_\chi)+2\beta_{\chi}\right]; ~ \text{electric},\\
\end{cases}
\end{split}
\label{eq:xsec_sch}
\eea
and $t$-channel $\chi \bar{\chi} \to \gamma \gamma$ annihilation cross-section is expressed as
\bea
\sigma_{\gamma\gamma}=\frac{s}{3\pi\Lambda^4(1-\beta_{\chi})}\left[1+12\beta_{\chi}-72\beta_{\chi}^2+32\beta_{\chi}^3-24\beta_{\chi}^2\log\left(\frac{1-2\beta_{\chi}}{2\beta_{\chi}}\right)\right];~ \text{magnetic/electric},~
\label{eq:xsec_tch}
\eea
where $N_f=1(3)$ is the color degrees of freedom for charged leptons (quarks), $\alpha_e$ is the fine structure constant and $\beta_{i}=m_i^2/s$. The correct DM relic abundance is evaluated when
\beq
m_{\chi} Y_0=\frac{\Omega_{\text{DM}}h^2 \rho_c}{s_0h^2} \simeq 4.3 \times 10^{-10}~\text{GeV},
\eeq
with $\rho_c \simeq 1.05 \times 10^{-5}\, h^2\, \mathrm{GeV/cm^3}$ is the critical energy density of the universe, $\Omega_{\mathrm{DM}} h^2 \simeq 0.120$ is the observed DM relic abundance as measured by the Planck collaboration \cite{Planck:2018vyg}, and $s_0\simeq 2.69 \times 10^{-3}$ cm$^{-3}$ is the present day entropy density. Following the condition $\Gamma_{\chi}\equiv n_{\text{eq}}(T_{\text{FO}})\langle \sigma v \rangle=\mathcal{H}(T_{\text{FO}})$, the freeze-out temperature during radiation domination is evaluated as
\beq
    T_{\text{FO}}=-\frac{2m_{\chi}}{\mathcal{W}_{-1}^{}\left[-\frac{8\pi^5}{45}\frac{g_{\rho}^{}}{g^2_{}}\frac{1}{\left(M_{Pl}^{}m_{\chi}^{}\langle\sigma v\rangle\right)^2}\right]}\,,
\eeq
where $\mathcal{W}_{-1}$ is the $-1$ brunch of the Lambert function.

When freeze-out takes place during the reheating, the SM entropy is continuously injected through inflaton decays and is therefore not conserved. In this case, it is more convenient to rewrite the Boltzmann equation in eq.~\eqref{eq:beq} in the following form
\beq
\frac{dN}{da}=-\frac{s\langle \sigma v \rangle}{\mathcal{H}a^4}(N^2-N^2_{\text{eq}}),
\label{eq:beq_FO_a}
\eeq
with $N_{\chi}=n_{\chi}\times a^3$ and $N_{\text{eq}}=n_{\text{eq}}\times a^3$. When freeze-out occurs out of equilibrium during reheating, the comoving DM number density following eq.~\eqref{eq:beq_FO_a} yields,
\beq
\frac{N_{\chi}}{a_{\text{RH}}^3}\simeq
\frac{3\mathcal{H}_{\text{RH}}}{2\langle \sigma v\rangle}
	\left(\frac{a_{\text{FO}}}{a_{\text{RH}}}\right)^{3/2},
\label{eq:FO_RH_ndm}
\eeq
where $a_{\text{FO}}=a_{\text{RH}}\left(T_{\text{RH}}/T_{\text{RO}}\right)^{3/8}$ is the scale factor at DM freeze-out and $\mathcal{H}_{\text{RH}}\equiv \mathcal{H}(T_{\text{RH}})=(8\pi^2/3)\sqrt{g_{s}^{*}(T_{\text{RH}})/10}\,(T_{\text{RH}}^2/M_P)$.

\begin{figure}[t]
	\centering
	\includegraphics[width=0.49\linewidth]{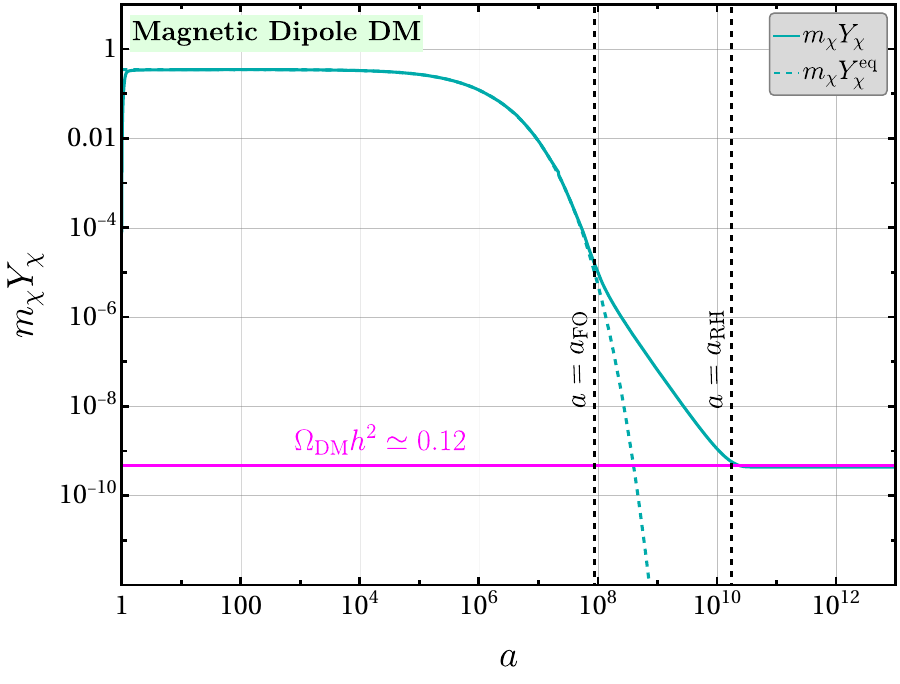}~~
	\includegraphics[width=0.49\linewidth]{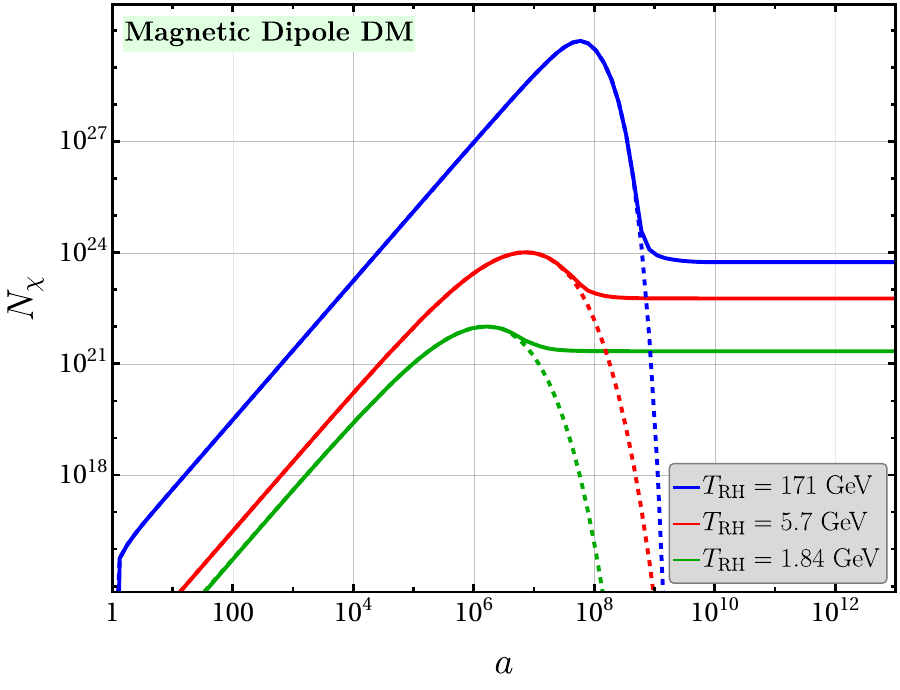}
	\caption{Variation of WIMP DM yield (left) and comoving DM number (right) as with the scale factor for magnetic dipole DM. We take $\{m_{\chi},~\Lambda,~T_{\text{RH}}\}=\{115,~5 \times 10^4,~2.5\}$ GeV for left panel and $\{m_{\chi},~\Lambda\}=\{115,~5 \times 10^4\}$ GeV for right panel.}
	\label{fig:DM_WIMP}
\end{figure}

Figure~\ref{fig:DM_WIMP} shows the DM yield and comoving number evolution obtained by numerically integrating the coupled Boltzmann equations \eqref{eq:inf_rad} and \eqref{eq:beq_FO_a} alongside the Friedmann equation \eqref{eq:hub_eq} for magnetic dipole DM\footnote{Electric dipole DM also exhibits similar behavior.}. As evident from the left panel of figure~\ref{fig:DM_WIMP}, the DM comoving number density undergoes dilution following chemical decoupling, persisting until the Universe reaches the reheating temperature. This behavior stands in stark contrast to the standard freeze-out scenario in the radiation-dominated era, where the yield remains constant after freeze-out. The dilution is a direct consequence of entropy injection from the decaying inflaton field during reheating. In the right panel, we present the evolution of comoving DM number density for different $T_{\text{RH}}$. The earlier the freeze-out, the longer the window over which entropy injection can dilute the DM yield, ultimately suppressing the DM relic abundance. This can be understood from the lower $T_{\text{RH}}$, as the prolonged reheating phase leads to longer dilution and a correspondingly smaller relic density.

\subsection{Freeze-in}
\label{sec:fi}
In the case of freeze-in mechanism, the DM can never reach thermal equilibrium, i.e. $n_{\chi}<<n_{\text{eq}}$, therefore the DM number density governed by the Boltzmann equation
\beq
\frac{dn_{\chi}}{dt}+3\mathcal{H}n_{\phi}^{}=\mathcal{C}_{\text{int}}.
\label{eq:beq_fi}
\eeq
Here, $\mathcal{C}_{\text{int}}$ denotes the DM interaction rate density which is parameterized as \cite{Elahi:2014fsa}
\beq
\mathcal{C}_{\text{int}}=\frac{T^{k+6}}{\Lambda^{k+2}},
\eeq
with $k=2(d-5)$ for the an effective DM-SM operator of dimension $d$ ($d \ge 5$) assuming $m_{\chi} \ll T$ . In practice, we take the exact expression of $\langle \sigma v \rangle$ in $\mathcal{C}_{\text{int}}$ to solve Boltzmann equation numerically. As stated before, during reheating, the SM entropy is not conserved, therefore we express eq.~\eqref{eq:beq_fi} in terms of the comoving number density as
\beq
\frac{dN}{da}=\frac{a^2 \mathcal{C}_{\text{int}}}{\mathcal{H}}.
\label{eq:beqa_fi}
\eeq
Depending on the DM mass, two distinct cases may arise: (i) for $m_{\chi} \ll T_{\text{RH}}$, the DM is predominantly produced at the end of reheating, with the DM comoving number density
\beq
\frac{N_{\chi}(a_{\text{RH}})}{a_{\text{RH}}^3} \simeq \frac{4 g_{\chi}^2 \zeta(3)^2 \kappa}{9\pi^4} \frac{T_{\text{RH}}^6}{\Lambda_{\text{eff}}^2  \mathcal{H}_{\text{RH}}} \left[1-\left(\frac{a_I}{a_{\text{RH}}}\right)^{9/4}\right],
\label{eq:ndm_FI_RH1}
\eeq
where the relativistic equilibrium number density is given by $n_{\rm eq} = \kappa g_{\chi}\zeta(3) T^3/\pi^2$
with $\kappa=3/4$ corresponding to fermionic DM. In addition, we assume a vanishingly small initial DM abundance at the onset of reheating $N_{\chi}(a_I)\approx 0$. (ii) On the other hand, for the condition $T_{\text{RH}}<m_{\chi} < T_{\text{max}}$, DM is produced during reheating. The comoving number density in this can be evaluated as
\beq
\frac{N_{\chi}(a_{\text{RH}})}{a_{\text{RH}}^3} \simeq \frac{4 g_{\chi}^2 \zeta(3)^2 \kappa}{9\pi^4} \frac{T_{\text{RH}}^{12}}{\Lambda_{\text{eff}}^2  \mathcal{H}_{\text{RH}}m_{\chi}^6} \left[1-\left(\frac{a_I}{a_{\text{DM}}}\right)^{9/4}\right],
\label{eq:ndm_FI_RH2}
\eeq
with $a_{\text{DM}}=a_{\text{RH}}(T_{\text{RH}}/m_{\chi})^{3/8}$. From eqs.~\eqref{eq:ndm_FI_RH1} and \eqref{eq:ndm_FI_RH2}, it is evident that the DM number density strongly dependent on $T_{\text{RH}}$, indicating the UV freeze-in nature of DM production.
\begin{figure}[t]
	\centering
	\includegraphics[width=0.49\linewidth]{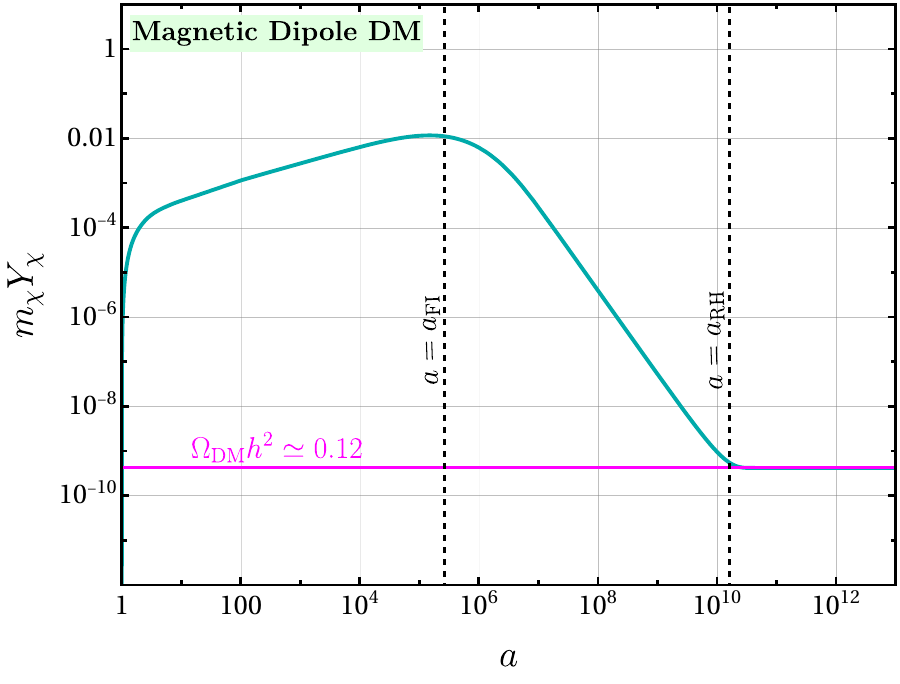}~~
	\includegraphics[width=0.49\linewidth]{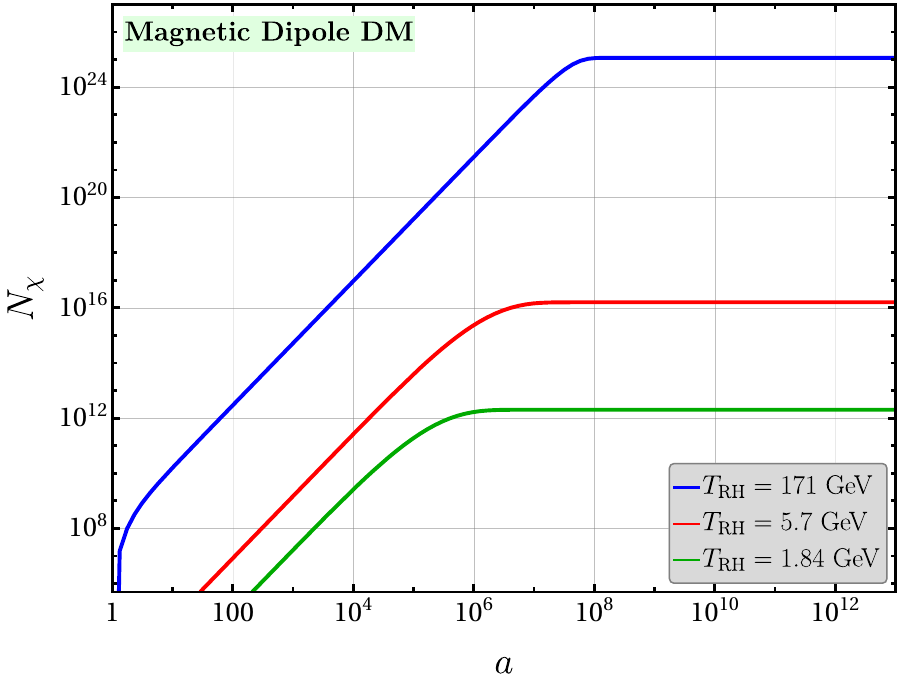}
	\caption{Variation of FIMP DM yield (left) and comoving DM number (right) as with the scale factor for magnetic dipole DM. We take $\{m_{\chi},~\Lambda,~T_{\text{RH}}\}=\{25,~5 \times 10^{11},~2.5\}$ GeV for left panel and $\{m_{\chi},~\Lambda\}=\{25,~5 \times 10^{11}\}$ GeV for right panel.}
	\label{fig:DM_FIMP}
\end{figure}
Figure~\ref{fig:DM_WIMP} illustrates the evolution of the DM yield and the comoving DM number in the freeze-in scenario for magnetic dipole DM. The left panel clearly demonstrates the effect of entropy dilution on the DM yield. The right panel shows that, similar to the WIMP scenario, a lower  $T_{\rm RH}$ leads to a smaller comoving DM number due to the extended reheating period, during which entropy production dilutes the DM abundance. 

We note that DM may also be generated through the direct decay of the inflaton. In that case, the Boltzmann equation in eq.~\eqref{eq:beq} acquires an additional source term proportional to $\mathcal{B}_{\rm DM}\Gamma_\phi n_\phi$, where $\mathcal{B}_{\rm DM}$ denotes the branching fraction of the inflaton into a pair of DM particles, and $n_\phi$ is the inflaton number density. In this work, however, our primary objective is to investigate how $\Gamma_{\phi}$ rate influences the production of DM. We therefore neglect the contribution arising from direct inflaton decays. A comprehensive discussion of DM production considering direct decay from inflaton can be found in Refs.~\cite{Harigaya:2014waa,Harigaya:2019tzu,Drees:2021lbm,Drees:2022vvn,Mukaida:2022bbo}.
  
\section{Constraints}
\label{sec:cons}
\subsection{Direct searches}
Direct detection experiments search for DM scattering off electrons/nucleons, producing detectable recoil signals of the target particle in terrestrial detectors. In the DM mass range of $1-10$ GeV, both DM-electron and DM-nucleon scattering processes play important roles in constraining the viable parameter space. For dipole interactions, the DM-electron scattering process is governed by the $t$-channel diagram and corresponding cross-section is given by \cite{Essig:2011nj}
\bea
\overline{\sigma}_{\chi e}\equiv \dfrac{\mu_{\rm \chi e}^2}{16\pi m_{\chi}^2 m_e^2}\overline{|\mathcal{M}_{\chi e}(q)|^2}\bigg|_{|{\bf q}|^2=\alpha^2m_e^2}.
\eea
Existing constraints on DM-electron scattering cross-section from experiments such as the XENON1T~\cite{XENON:2019gfn} and PandaX-4T~\cite{PandaX:2022xqx} already impose strong bound on the parameter space, particularly for the DM masses in the range of 1 to 10 GeV. In this mass range, the most stringent bound is provided by the PandaX-4T experiment, which constrains the scale $\Lambda$ to be of the order of $10^{4}$ TeV. Next-generation experiments such as Oscura~\cite{Oscura:2023qik} is expected to improve the sensitivity to the DM-electron scattering cross-section by approximately two orders of magnitude.  Consequently, for electromagnetic dipole DM with masses up to 10 GeV, Oscura will be able to probe the cutoff scale $\Lambda$ up to about six times higher than the current limit set by the PandaX-4T experiment. In the context of DM-nucleon scattering, the PandaX-4T collaboration~\cite{PandaX-4T:2021bab} also provides stringent constraints in this mass range, probing the cutoff scale up to approximately $9 \times 10^{6}~(6 \times 10^{9})$ GeV for magnetic (electric) dipole DM at a DM mass up to 10 GeV.

For DM masses above 10 GeV and extending up to the multi-TeV scale, constraints from DM-nucleon scattering become relevant in shaping the viable parameter space. Over the years, a number of direct detection experiments have been probing nuclear recoils induced by DM interactions. In this context, experiments such as PandaX-4T~\cite{PandaX:2024qfu}, LUX-ZEPLIN (LZ)~\cite{LZ:2024zvo}, and XENONnT~\cite{XENON:2025vwd} have conducted dedicated searches for DM via DM-nucleon scattering. Among these, the LZ experiment currently provides the most stringent limit on the DM-nucleon scattering cross-section for DM mass range from 10 GeV up to $10^4$ GeV. The proposed XLZD experiment \cite{XLZD:2024nsu} is expected to achieve a sensitivity to the DM--nucleon scattering cross section that is approximately one order of magnitude stronger than that of the current LZ experiment.

The direct search constraints presented in figure~\ref{fig:lam_mass}. For magnetic (electric) dipole DM, the stringent constraint on the scale $\Lambda$ is approximately $4.5 \times 10^7~(4 \times 10^{10})$ GeV for DM mass approximately 35 GeV. Therefore, the direct search constraint on the scale $\Lambda$ for electric dipole DM is more stringent than the magnetic dipole DM. This can be understood by noting that, in the non-relativistic limit, magnetic dipole interactions contribute via spin-dependent scattering with nucleons. In contrast, electric dipole DM gives rise to both spin-independent and spin-dependent interactions, leading to an overall enhancement of the scattering rate and consequently tighter experimental constraints. For electromagnetic dipole DM, the projected XLZD sensitivity extends the exclusion reach on the scale $\Lambda$ by approximately a factor of three compared to the current LZ constraints.

\subsection{Indirect searches}

The Fermi-LAT \cite{Fermi-LAT:2015kyq} and H.E.S.S. \cite{HESS:2020zwn} collaborations have accumulated more than a decade of observations in the direction of the Galactic Center, leading to constraints on DM parameter space over a wide mass range. The Fermi-LAT observation provides the strongest bounds for DM masses up to 300 GeV. In the intermediate regime, $300\,\mathrm{GeV} \lesssim m_\chi \lesssim 10\,\mathrm{TeV}$, both Fermi-LAT and H.E.S.S. yield comparable sensitivities. For heavier DM masses above 10 TeV, the constraints are dominated by the H.E.S.S. observations. In terms of the effective scale, at $m_\chi \sim 300\,\mathrm{GeV}$ the constraint on $\Lambda$ is approximately $10\,\mathrm{TeV}$, while H.E.S.S. strengthens this limit to $\Lambda \sim 30$ TeV for DM mass extending up to 40 TeV. These constraints are essentially similar for both magnetic and electric dipole DM, since the annihilation cross section $\chi \bar{\chi} \to \gamma \gamma$ for both the cases are same.

\subsection{Collider searches}

We now discuss the collider constraints relevant for dipole DM searches. Invisible particle searches using a mono-$\gamma$ plus missing energy signature have been performed at the LEP \cite{DELPHI:2008uka}. The results were recast in terms of the photon-energy distribution \cite{Fox:2011fx}, and subsequently reinterpreted to constrain dipole DM scenarios \cite{Fortin:2011hv}. The corresponding bound on $\Lambda$ is approximately 670 GeV for DM masses up to 100 GeV, for both magnetic and electric dipole DM. On the other hand, results from the CMS collaboration at the LHC based on the di-jet plus missing energy channel \cite{CMS:2017zts} have also been used to constrain dipole DM \cite{Arina:2020mxo}. This analysis shows that LHC data place a bound of $\Lambda \sim 1$ TeV (400 GeV) for DM masses up to $\sim 1~\mathrm{TeV}$ for magnetic (electric) Dipole DM, while projections for the HL-LHC indicate sensitivity up to approximately $\Lambda \sim 1.2~\mathrm{TeV}$ (3 TeV) for magnetic (electric) dipole DM.

\section{Neutron star heating}
\label{sec:ns_heating}
Inside the DM halo, the DM particles are typically non-relativistic. However, when they approach a massive compact object such as a neutron star, the strong gravitational potential can accelerate them to velocities of order $\mathcal{O}(0.3c)$. As these particles fall into the neutron star, they undergo scattering with nucleons, losing enough energy to become gravitationally bound. Once captured, DM accumulates inside the star and, over relevant timescales, efficiently transfers its kinetic energy to the stellar medium. Consequently, the initial kinetic energy of the infalling DM is effectively converted into heat, contributing to the thermal evolution of the neutron star. For the DM mass in the range $1-10^{6}$ GeV\footnote{We note that for DM mass below 1 GeV, Pauli blocking effects become significant \cite{Joglekar:2020liw,Bell:2020jou}, while for masses above $10^{6}$ GeV, multiple scatterings are required for the particle to lose sufficient kinetic energy for capture \cite{Bramante:2017xlb,Dasgupta:2019juq}. Since such effects lie beyond the scope of this work, we restrict our analysis to this intermediate mass range.}, capture inside a neutron star can typically be achieved through a single scattering event, as one interaction is sufficient to reduce the particle’s kinetic energy below the escape threshold. Under this framework, the entire initial kinetic energy of the DM that transferred to the heating, is expressed as \cite{Bell:2023ysh}
\beq
T_{\text{eff}}=2410 f_{\text{cap}}^{1/4}\left(\frac{\rho_{\text{DM}}}{0.4~\rm{GeV/cm^{-3}}}\right)^{1/4},
\label{eq:temp}
\eeq
where $\rho_{\text{DM}}$ denotes the DM density in the vicinity of the neutron star, taken to be the local value $0.4\,\mathrm{GeV/cm^3}$, and the velocity of the neutron star is assumed to be the same as that of the Sun.  $f_{\rm cap}$ represents the fraction of DM particles captured within the neutron star and is given by
\beq
f_{\rm cap} \sim {\rm Min} \left[ \sum_i \frac{\sigma_{\chi i}}{\sigma_{ {\rm th},i }}, 1  \right],
\label{eq:f_cap}	
\eeq
where $\sigma_{\chi i}$ indicates the DM scattering cross-section with a target species $i$, and $\sigma_{{\rm th},i}$ is the corresponding threshold cross section for capture inside the neutron star. It is defined as $\sigma_{{\rm th},i} = \pi R_{\rm NS}^2 / N_i$, where $m_i$ and $N_i$ represent the mass of the target particle and the total number of such scattering targets within the neutron star, respectively.

In order to evaluate the capture fraction defined in eq.~\eqref{eq:f_cap}, we include contributions from both protons and neutrons. Protons couple to the photon at tree level through the electromagnetic interaction, whereas neutrons, being electrically neutral, interact primarily via a magnetic dipole term that arises at loop level. For the dipole DM scenario considered here, these tree-level proton contributions and loop-induced neutron contributions are of comparable magnitude. Adopting a conservative approach, we assume a proton fraction of $1\%$ within the neutron star \cite{Sedrakian:2006mq}. In the relativistic limit, the differential scattering cross-sections of magnetic dipole DM with protons and neutrons by
\bea
\begin{split}
 \frac{d \sigma_{ \chi p \rightarrow \chi p }}{ d \cos \theta }  & = \frac{1}{ 32 \pi s } \times \frac{64 \pi \alpha_e}{\Lambda^2 t} \left[ 2m_{\chi}^2 \left( s+t \right) - m_p^4 - m_{\chi}^4 + m_p^2 \left( 2m_{\chi}^2 + 2s + t \right) - s \left( s+t \right) \right],	\\
 \frac{d \sigma_{ \chi n \rightarrow \chi n }}{ d \cos \theta } & = \frac{1}{ 32 \pi s } \times \frac{4 \mu_n^2}{\Lambda^2} \left[ 4 \left( m_n^4 + m_{\chi}^4 \right) - 8 m_{\chi}^2 s  + 8 m_n^2 \left( 3 m_{\chi}^2 - s \right) + \left( 2 s + t \right)^2 \right],
\end{split}
\label{eq:diff_mdm}
\eea
while for electric dipole DM, the expressions are
\bea
\begin{split}
\frac{d \sigma_{ \chi p \rightarrow \chi p }}{ d \cos \theta } & = \frac{1}{ 32 \pi s } \times \frac{64 \pi \alpha_e}{\Lambda^2 t} \left[ t \left( m_p^2-s \right) - \left( m_p^2 + m_{\chi}^2 -s \right)^2 \right],\\
\frac{d \sigma_{ \chi n \rightarrow \chi n }}{ d \cos \theta } & = \frac{1}{ 32 \pi s } \times \frac{4 \mu_n^2}{\Lambda^2} \left[ 4 m_n^4 - 8 m_n^2 \left( m_{\chi}^2 + s \right) + \left( 2s + t - 2 m_{\chi}^2 \right)^2 \right].
\end{split}
\label{eq:diff_edm}
\eea
In the above expressions of the differential cross sections, $m_{p(n)}$ is the mass of the proton, $s$ and $t$ are the Mandelstam variables, $\cos\theta$ is the cosine of the scattering angle in the CM frame, and $\mu_{n}$ is the neutron dipole moment. Sufficiently old neutron stars can cool to temperatures of order $\mathcal{O}(1000\ \rm K)$, 
rendering them within the observational reach of next-generation infrared telescopes such as the James Webb Space Telescope (JWST)~\cite{Gardner:2006ky}, the Thirty Meter Telescope (TMT)~\cite{Crampton:2008gx}, and the European Extremely Large Telescope (E-ELT)~\cite{Maiolino:2013bsa}. As demonstrated in Ref.~\cite{Chatterjee:2022dhp}, JWST is capable of detecting neutron stars with surface temperatures $\gtrsim 2400\ \rm K$ within 
the local vicinity of the Solar neighborhood. Motivated by this observational threshold, we derive projected sensitivities on the $\Lambda-m_{\chi}$ plane by requiring that the combined heating of the neutron star, arising from both dark kinetic heating and DM annihilation, sustains a surface temperature of $2400\ \rm K$. We find that JWST can probe the cut-off scale approximately up to $4~(6) \times 10^{8}$ GeV for magnetic (electric) dipole DM, across the mass range  $1-10^6$ GeV considered in this analysis.

\begin{figure}[t]
	\centering
	\includegraphics[height=6cm,width=8cm]{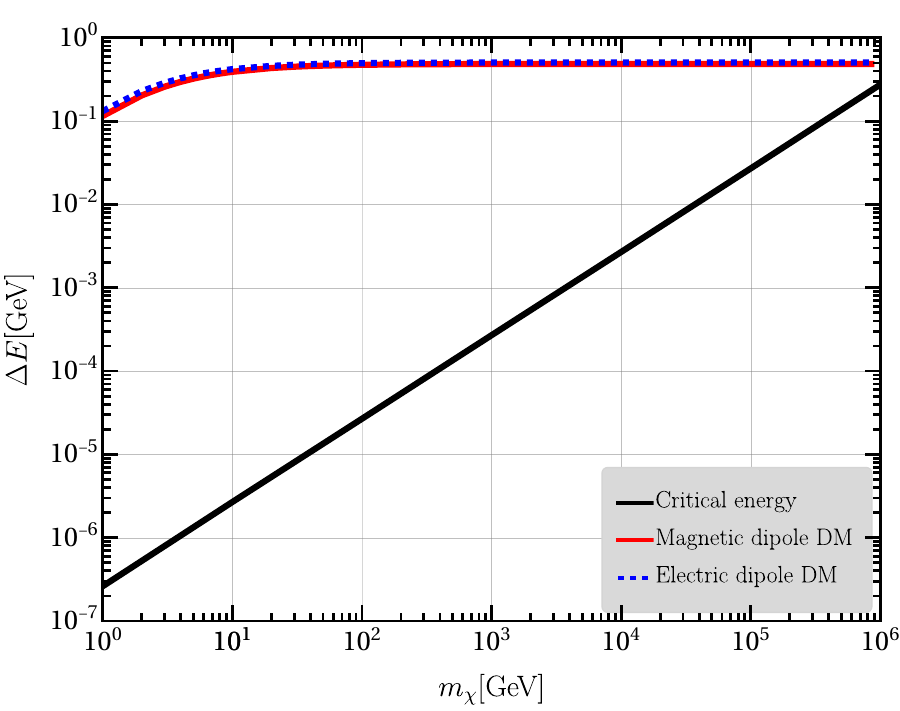}
	\caption{Variation of the average energy transfer per single scattering event inside a neutron star as a function of DM mass for different DM scenarios. The black solid curve indicates the critical energy loss required for successful capture of DM particles. The red solid (blue dashed) curves represent the average energy loss for the magnetic dipole and electric dipole DM cases, respectively.}
	\label{fig:en_trans}
\end{figure}

\subsection{Energy transfer for a single scattering}
Here, we discuss the average energy transferred by a DM particle in a single scattering event inside a neutron star. The mean energy transfer in a typical DM--nucleon scattering process is given by \cite{Bell:2018pkk,Bell:2019pyc}
\begin{equation}
\langle \Delta E \rangle  = \dfrac{\left( 1-B \right) m_{\chi} \zeta_p}{B + 2 \sqrt{B} \zeta_p + B \zeta_p^2}  \times
		\dfrac{\int_{-1}^{1} d \cos \theta  \left( 1 - \cos \theta \right) \left( \frac{d \sigma}{d \cos \theta} \right)}
		{\int_{-1}^{1} d \cos \theta \left( \frac{d \sigma}{d \cos\theta}\right)},
		\label{eq:en_trans}
\end{equation}
where $\zeta_p = m_{\chi}/m_p$, and
$B = \left(1 - 2 G M_{\rm NS}/R_{\rm NS} \right) \simeq 0.56$ for a typical neutron star configuration considered in this work. For successful capture, the energy transferred from the incoming DM particle to the stellar constituent must exceed a critical value, $E_{\rm cr}(m_{\chi})$, corresponding to the ambient kinetic energy of the DM particle \cite{Raj:2017wrv}. In figure~\ref{fig:en_trans}, the critical energy loss is shown by the black curve, assuming an average Galactic halo DM velocity of $\sim 220$ km/sec. The red curve (blue dashed curve) in figure~\ref{fig:en_trans} represents the average energy loss in the magnetic (electric) dipole DM scenario, obtained from eq.~\eqref{eq:en_trans} using the expression differential scattering cross section given in eq.~\eqref{eq:diff_mdm} (eq.~\eqref{eq:diff_edm}).
We find that across the DM mass range  considered in this work, the average energy transfer exceeds the critical energy loss required for DM capture due to momentum dependent interactions of electromagnetic dipole DM. 
\section{Results}
\label{sec:results}
\begin{figure}[t]
	\centering
	\includegraphics[height=10.96cm,width=12.5cm]{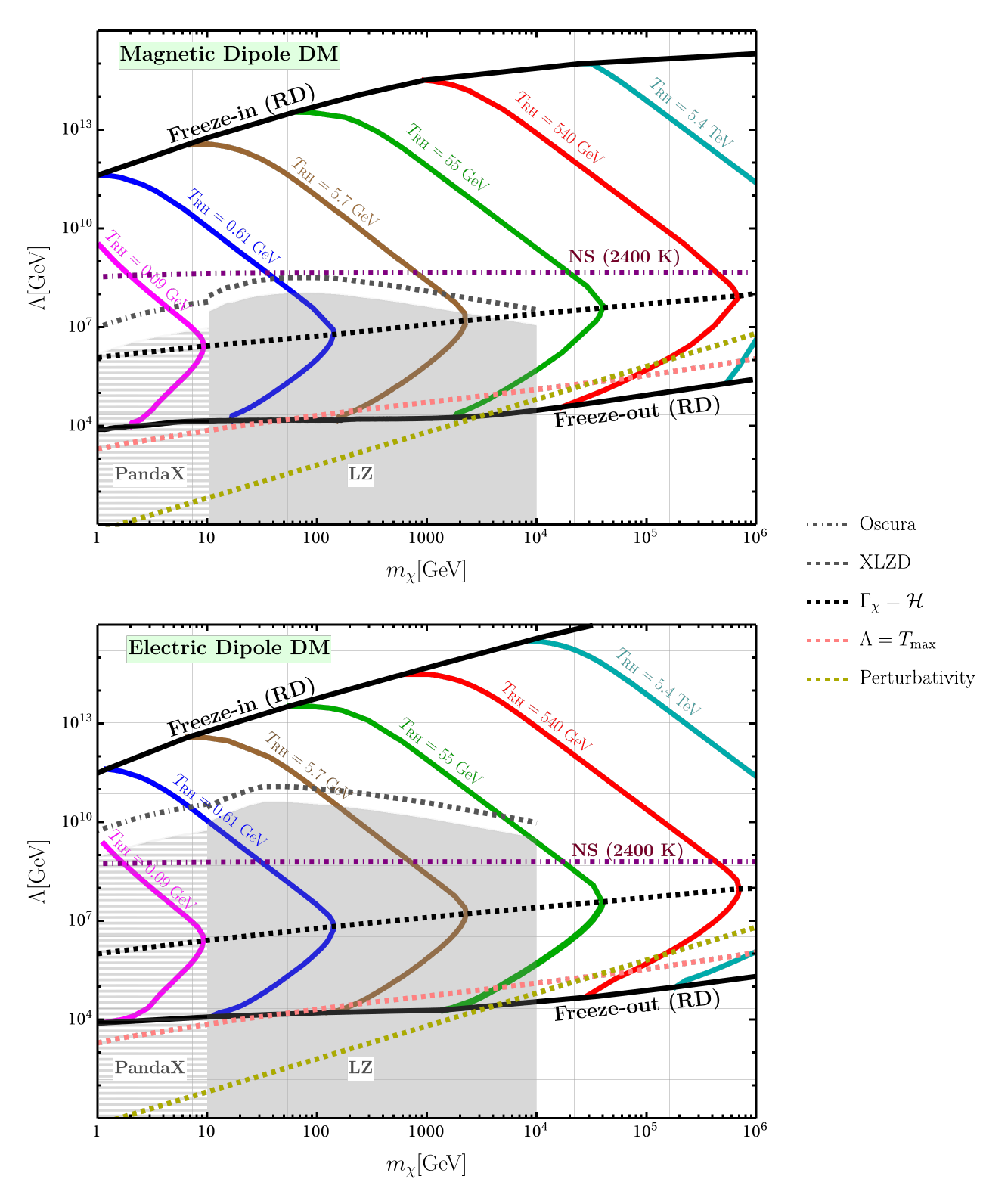}
	\caption{Parameter space for dipole DM. In each panel, the black solid curve represents the relic density allowed parameter space in the standard radiation-dominated (RD) era. The colored solid curves correspond to the relic density allowed regions for different reheating temperatures. The black dashed line indicates the boundary above which thermal equilibrium is not attained. The pink dashed line denotes the condition $\Lambda = T_{\rm max}$, while the mustard dashed line separates the region below where the perturbative expansion breaks down~\cite{Griest:1989wd}. The hatched and shaded regions are excluded by constraints from DM-nucleon scattering experiments~\cite{Garcia:2020eof}. The purple dot-dashed curve shows the projected sensitivity of JWST from neutron star heating corresponding to a surface temperature of 2400 K~\cite{Chatterjee:2022dhp}. The top panel corresponds to electric dipole DM, whereas the bottom panel represents magnetic dipole DM.}
	\label{fig:lam_mass}
\end{figure}

The evolution of the cosmological background is characterized by  $\mathcal{H}_I$ and $\Gamma_\phi$, or equivalently by $T_{\max}$ and $T_{\rm rh}$. In this work, we adopt $\mathcal{H}_I = 0.1$ GeV in our analysis which is within the current observational limit mentioned above. The viable parameter spaces for magnetic (top panel) and electric (bottom panel) dipole DM in the $\Lambda-m_{\chi}$ plane are shown in figure~\ref{fig:lam_mass}. We take six values of the inflaton width, $\Gamma_{\phi} = 10^{-20},~10^{-18},~ 10^{-16},~10^{-14},~10^{-12},~\text{and}~10^{-10}$~GeV, which correspond to $T_{\text{RH}} \simeq$ 0.09, 0.61, 5.71, 55.28, 540 and 5400 GeV, respectively (and to $T_{\text{max}} \simeq$ 2.8, 9.12, 28.85, 91.24, 288, and 912 TeV, respectively).
In each plot, the top (bottom) black solid curve corresponds to FIMP (WIMP) scenario that yield the observed relic density, where freeze-in (freeze-out) occurs during radiation domination. In these scenarios, the DM relic density in insensitive to the $T_{\text{RH}}$ of the Universe. The black dashed line represents $\Gamma_\chi=\mathcal{H}$, the minimum value of $\Lambda$ for which the chemical equilibrium is achieved. The region above the black dashed line corresponds to scenarios in which DM never attains chemical equilibrium with the thermal bath and is therefore produced through non-thermal processes. In contrast, the region enclosed between the black dashed and bottom black solid lines represents scenarios with sufficiently large DM annihilation cross-sections, causing DM to remain in chemical equilibrium for an extended period and freeze-out relatively late. As a consequence, the resulting relic abundance becomes under-abundant ($\Omega_{\rm DM} h^2 < 0.12$) during standard radiation domination. The region below bottom black solid line disallowed from the overproduction of WIMP ($\Omega_{\mathrm{DM}} h^2 > 0.12$). The top solid black curve dictates FIMP DM consistent with the observed relic density, in which DM production is maximized around $m_{\chi} \sim T_{\mathrm{FI}}$ in the radiation-dominated epoch. The region above the top solid black line is ruled out from the underabandent FIMP DM and the region between top solid black line and black dashed line ruled due to overproduction of the FIMP DM during standard radiation domination. 

If we consider DM production during reheating ($T_{\mathrm{RH}}<m_{\chi}<T_{\mathrm{max}}$), considering a specific reheating temperature $T_{\mathrm{RH}}$ induced by inflaton decay width $\Gamma_{\phi}$, this leads to interesting phenomenological consequences. In particular, regions of parameter space that are excluded in the standard radiation-dominated picture between the WIMP and FIMP regimes can become accessible. During reheating, For a fixed $T_{\mathrm{RH}}$, an increase in the DM mass leads to a higher freeze-out temperature, thereby widening the gap between $T_{\mathrm{RH}}$ and $T_{\mathrm{FO}}$. This larger gap enhances the entropy dilution effect, reducing the DM relic abundance. To counterbalance this reduction and recover the observed DM relic density, the DM annihilation cross-section must be decreased, which can be achieved by increasing $\Lambda$. In contrast, for the FIMP case, the DM yield scales with the production rate. Consequently, the entropy injection is counterbalanced by smaller values of $\Lambda$. It is worthwhile to note that a suitable reheating temperature enables a continuous transition between the disconnected WIMP and FIMP regimes.

In the standard radiation-domination, for both magnetic and electric dipole DM, the parameter region corresponding to DM masses in the range $1-10^{4}$ GeV is largely excluded by current direct detection constraints. The relic density allowed region for WIMP DM with masses between $10^{4}$ and $10^{6}$ GeV  lies predominantly in the non-perturbative regime, rendering the EFT description unreliable. On the other hand, in the FIMP scenario within radiation domination, the relic abundance is directly proportional to the thermally averaged annihilation cross section. Consequently, to reproduce the observed DM abundance requires comparatively large values of the cutoff scale, approximately $10^{12}-10^{15}~(10^{12}-10^{16})$ GeV for magnetic (electric) dipole DM in the context of DM mass range considered here. Such large scales remain far beyond the reach of current experimental searches. 

The injection of entropy during reheating allows a large $\Lambda$ to satisfy the observed relic density for WIMPs, which in principle helps evade current direct detection bounds. For electromagnetic dipole DM, however, direct search constraints are sufficiently strong, excluding the parameter space for both during reheating and standard radiation domination scenarios up to DM mass of $10^4$ GeV for each dipole case. On the other hand, a lower $\Lambda$ is required to satisfy the observed relic density for FIMPs, which could be probed in the future run of the direct search experiments for the DM mass up to $10^4$ GeV. For magnetic dipole DM, NS heating provides a powerful probe of the reheating-induced FIMP parameter space, covering a significant part of the allowed region over the entire DM mass range considered. In the WIMP scenario, NS heating becomes particularly effective for heavier DM, probing the DM mass $ \gtrsim 10^{4}$ GeV up to $10^{6}$ GeV, irrespective of whether the relic abundance is determined during reheating or the standard radiation-dominated era. On the other hand, for electric dipole DM, existing direct detection limits remain more stringent than the projected NS heating sensitivity for DM mass up to $10^{4}$ GeV. However, for heavier DM ($10^{4}~\mathrm{GeV} \lesssim m_\chi \lesssim 10^{6}~\mathrm{GeV}$), NS heating is capable of probing nearly the entire WIMP parameter space similar to the magnetic dipole DM case. During reheating, the consistency of the effective DM-SM interaction imposes $\Lambda > T_{\max}$ \cite{Garcia:2020eof}, as indicated by the pink dashed line in the plot. The perturbative limit using the condition $2e m_{\chi} \geq \Lambda$ \cite{Sigurdson:2004zp} is denoted by the mustard dashed line. The unitarity limit can be obtained from the $s$-channel annihilation cross-section given in eq.~\eqref{eq:xsec_sch} by imposing the condition $\sigma \leq 4\pi/m_{\chi}^2$ \cite{Griest:1989wd}. Since this bound is approximately one order of magnitude weaker than the perturbative limit, we do not display it in the plot to avoid overcrowding the figure\footnote{For the same reason, we do not display the constraints from indirect detection and collider searches, as they are entirely superseded by the more stringent constraints from direct detection across the parameter space of interest.}. 
\section{Conclusion}
\label{sec:conclude}

In this work, we have explored the phenomenology of electromagnetic dipole DM within the effective field theory framework. We have evaluated the DM relic abundance in both freeze-out and freeze-in scenarios, considering cosmological evolution during the standard radiation-dominated era as well as during reheating. We have further investigated the relevant constraints on DM parameter space arising from direct detection, indirect searches, and collider experiments. Our analysis shows that direct detection limits from DM-nucleon scattering currently provide the most stringent constraints on the parameter space. In the standard radiation-dominated freeze-out scenario, for both the DM cases, the relic density allowed parameter space is largely excluded by present direct detection bounds, whereas the viable freeze-in parameter space lies far beyond the reach of current experiments. During reheating phase, entropy injection from inflaton decay significantly modifies the viable parameter space. For freeze-out production during reheating, to satisfy the observed relic density, the required cut-off scale increases for a given reheating temperature. However, the direct search constraints are sufficiently strong to rule out the WIMP DM parameter space up to DM mass $10^4$ GeV. In contrast, the cut-off scale decreases in the freeze-in scenario due to the same entropy dilution effect and part of the parameter space can be probe via future direct search experiment.

We have also studied the prospects of probing magnetic and electric dipole DM through compact neutron star heating induced by DM capture. Motivated by the capability of JWST to detect neutron stars with surface temperatures around $2400\,\mathrm{K}$, we have estimated the corresponding sensitivity to the DM parameter space. For electric dipole DM, the projected JWST sensitivity is already surpassed by current direct detection constraints for DM mass up to $10^{4}$ GeV. However, in the case of magnetic dipole DM, a viable parameter region still survives within this mass range. Furthermore, for DM mass between $10^{4}$ and $10^{6}\,\mathrm{GeV}$, neutron star heating observations with JWST can play a crucial role in probing potential DM signatures for both WIMP and FIMP candidates, across standard and non-standard cosmological scenarios. Such observations could therefore offer valuable insight into the phenomenology of electromagnetic dipole DM interactions, spanning different production mechanisms and cosmological histories.
\section*{Acknowledgments}
The author is grateful to Dipankar Pradhan for his invaluable assistance with the numerical implementation and for many insightful theoretical clarifications. The author also thanks Debajit Bose for useful discussions.

\appendix

\section{UV completion of magnetic dipole operator}
\label{sec:uv_completion}
In this section, we discuss a possible UV complete model that generate magnetic dipole operator. The model is extended by SM with a $SU(2)_L$ complex scalar doublet $\phi$ and a fermion doublet $\psi$ with the Dirac fermion DM candidate $\chi$. The Lagrangian for the model is given by
\beq
\mathcal{L} = \left( D_{\mu} \phi \right)^{\dagger} \left( D^{\mu} \phi \right) - M_{S}^{2} \phi^{\dagger} \phi + \bar{\chi}  \left( i \slashed{\partial} - m_{\chi} \right) \chi - m_{\chi} \bar{\chi} \chi + \bar{\psi} \left( i \slashed{D} - M_{\psi} \right) \psi - m_{\psi} \bar{\psi} \psi 
+ \lambda \bar{\psi} \chi \phi + \text{H.c.},
\eeq
where SM gauge invariant covariant derivative is defined as $D_{\mu} = \partial_{\mu} - i g W_{\mu}^{a} T^{a} -  ig' B_{\mu}/2 $ with $T^{a} = \sigma^{a}/2$ being the generators of $SU(2)_L$ and $Y=1/2$ the hypercharge. $g$ and $g'$ are the $SU(2)_L$ and $U(1)_Y$ gauge couplings, respectively. After integrating out the heavy degrees of freedom, the one-loop matching onto the dipole operator is carried out with \texttt{Matchete}~\cite{Fuentes-Martin:2022jrf}, yielding the corresponding relation: 
\beq
\frac{C_M}{\Lambda}= \frac{e \lambda^2}{48 \pi^2 M_{\psi}} \left[\frac{2r^2(r^2-1-2\log r)}{(1-r^2)^2}\right],
\eeq
with $r=M_{\psi}/M_{\phi}$.
\begin{figure}[htb!]
	\centering
\begin{tikzpicture}
	\begin{feynman}
		\vertex(a){$\chi$};
		\vertex[ right =2cm  of a] (a1);
		\vertex[ right =4cm  of a] (a2);
		\vertex[ right =6cm of a] (a3){$\chi$};
		\vertex[ above right =1cm and 1cm of a1] (b);
		\vertex[ above right=1.2cm and 1.2cm of b] (b1){$\gamma$};
		\diagram*{
			(a) -- [ fermion, arrow size=1.5pt, ultra thick] (a1),
			(a2) -- [scalar ,style=black!50,edge label={\({\color{black} S} \)}, ultra thick]  (a1), 
			(a2)-- [fermion,arrow size=1.5pt, ultra thick] (a3),
			(a1) -- [fermion,quarter left, arrow size=1.5pt, style=black!50, edge label={${\color{black}\rm\psi}$}, ultra thick] (b),
			(b) -- [fermion, quarter left,arrow size=1.5pt, style=black!50, edge label={${\color{black}\rm\psi}$}, ultra thick] (a2),
			(b)-- [boson,arrow size=0.7pt,edge label={${\color{black}\rm}$},style=black, ultra thick] (b1)};
	\end{feynman}
\end{tikzpicture}
\caption{One loop Feynman diagram responsible for generating magnetic dipole operator.}
\end{figure}
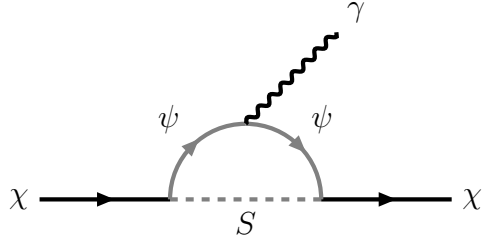	
\bibliographystyle{JHEP}
\bibliography{ref.bib}
\end{document}